\documentstyle[eqsecnum,aps,prb,multicol,epsf]{revtex}
\begin{document}

\def\top#1{\vskip #1\begin{picture}(290,80)(80,500)\thinlines \put(
65,500){\line( 1, 0){255}}\put(320,500){\line( 0, 1){
5}}\end{picture}}

\def\bottom#1{\vskip #1\begin{picture}(290,80)(80,500)\thinlines \put(
330,500){\line( 1, 0){255}}\put(330,500){\line( 0, -1){
5}}\end{picture}}

\tighten

\title{Possible Depinning Transition of a Single Flux Line \\
Near a Columnar Defect in Type II Superconductors}
\author{A. M. Ettouhami}
\address{Department of Physics, University of Colorado, Boulder, CO 80309}
\date{\today}
\maketitle

\begin{abstract}
We use the Feynman--Kleinert variational method [R.P. Feynman and H.~Kleinert, Phys. Rev. A {\bf 34}, 5080, 
(1986)] to calculate the partition function and effective pinning energy of a single flux line near a columnar pin 
in type II superconductors. It is found that there is a phase transition between a low temperature phase where 
the  flux line is localized near the columnar pin and where its internal modes fluctuations are bounded, and a 
high temperature, depinned phase where the flux line is delocalized and its internal fluctuations are
those of a free line.
\end{abstract}


\begin{multicols}{2}

\section{Introduction}

During the past decade, the static and dynamic properties of flux lines in high temperature 
superconductors (HTSC) in the presence of a random array of pinning centers have been the subject of a 
considerable amount of work (for recent reviews, see Blatter et {\em al.}\cite{Blatter-et-al}, and 
Nattermann and Scheidl\cite{Nattermann-Scheidl}). 
Among the various types of pinning phenomena which have been investigated, 
a particularly promising one is the relatively strong pinning which results from the interaction of flux lines 
with columnar defects\cite{Civale,Konczykowski}. The latter can be viewed as long, parallel damage tracks  
of diameter $d_0\approx 50-70 \AA$, produced by heavy ion (e.g. Sn, Pb) irradiation at high ($\sim$ GeV) energies.
Such columnar defects were found experimentally to lead to a significant enhancement of the irreversibility line, 
with critical currents increasing by as much as three orders of magnitude at $T=77 \,K$ in thallium-based 
compounds\cite{Budhani-et-al}. On the theoretical side, much of our present understanding of the pinning of flux 
lines by these columnar defects originates in the work of Nelson and Vinokur\cite{Nelson-Vinokur}, who used 
the boson analogy\cite{Nelson,Nelson-Seung} to study the physics of individual flux lines 
as well as flux line lattices and liquids in the presence of columnar pins.

In the present paper, we make an attempt at describing the physics of a single flux line-columnar 
pin pair, without resorting to the boson formulation. 
The flux line will be parametrized by the three-dimensional vector ${\bf R}(z)=({\bf r}(z),z)$, where the 
two-dimensional vector ${\bf r}(z)$ describes the position of the flux line element at height $z$ in the $(x,y)$ 
plane. We assimilate the flux line to an elastic string whose internal degrees of freedom can be described by the 
Hamiltonian\cite{Nelson-Vinokur}
\begin{eqnarray}
H_{el} = \int_0^L dz\,\,\frac{1}{2}\,\tilde\varepsilon_1\,\Big(\frac{d{\bf r}}{dz}\Big)^2
\label{el-H-1}
\end{eqnarray}
where $L$ is the sample thickness in the $z$ direction and 
$\tilde\varepsilon_1$ is the tilt energy per unit length of the flux line. For a superconductor with 
uniaxial symmetry, assuming that the external magnetic field ${\bf H}||{\bf c}$,  
$\tilde\varepsilon_1$ can be approximated by $\simeq\varepsilon^2\varepsilon_0\ln\kappa$, where 
$\varepsilon=\lambda_{ab}/\lambda_c$ is the ratio of the London penetration depths along the $(ab)$ plane and the 
${\bf c}$ axis, $\varepsilon_0=(\phi_0/4\pi\lambda_{ab})^2$ ($\phi_0=hc/2e$ is the flux quantum), and 
$\kappa=\lambda_{ab}/\xi_{ab}$ is the ratio of 
the London penetration depth to the coherence length in the $(ab)$ plane.

The total Hamiltonian of our system of one single flux line in the presence of a columnar pin is 
then given by \begin{eqnarray} 
H = \int_0^L dz\,\big\{\frac{1}{2}\,\tilde\varepsilon_1\,\Big(\frac{d{\bf r}}{dz}\Big)^2
+ V({\bf r}(z)) \big\}
\end{eqnarray}
where the $z$-independent function $V({\bf r})$ denotes the pinning potential due to the columnar defect.
$V({\bf r})$ will be taken as a very localized function, $V({\bf r})\approx -U_0$ for $|{\bf r}| \leq b_0$ 
and $V({\bf r})\approx 0$ for $|{\bf r}| > b_0$, with\cite{Nelson-Vinokur} 
$b_0=\mbox{max}(d_0/2,\sqrt{2}\xi_{ab})$ and $U_0\approx\varepsilon_0\ln[1+(d_0/2\sqrt{2}\xi_{ab})^2]/2$. 
If we take as the origin of free energies the free energy of a free flux line ({\em i.e.} a flux line far from the 
columnar pin), 
then the partition function of our system can be written as the {\em normalized} path 
integral\cite{Nelson-Vinokur}
\begin{equation}
Z = \frac{\int{\cal D}{\bf r}(z)\;\exp\Big\{
- \beta\int_0^L dz\,\big[\frac{\tilde\varepsilon_1}{2}\Big(\frac{d{\bf r}}{dz}\Big)^2  + 
V\big({\bf r}(z)\big) \big]
\Big\} }
{\int{\cal D}{\bf r}(z)\;\exp\Big\{
- \beta\int_0^L dz\;\frac{\tilde\varepsilon_1}{2}\Big(\frac{d{\bf r}}{dz}\Big)^2 
\Big\}}
\label{normalized-Z}
\end{equation}
with $\beta=1/k_BT$ the inverse temperature ($k_B$ is Boltzmann's constant).
In its simplest form, which happens also to be the form relevant to our present problem, the boson analogy
consists in exploiting the formal correspondence between the above, normalized partition function, 
and the partition function of a quantum particle of mass $m$ in imaginary 
time\cite{Feynman-Hibbs,Feynman,Negele-Orland}($\equiv\tau$), as can be seen by using the 
replacements\cite{Nelson-Seung} $z\to\tau$,
$k_BT\to\hbar$, $\tilde\varepsilon_1\to m$ and $L\to \beta\hbar$ ($\hbar$ is Planck's constant $h$ 
divided by $2\pi$).
This correspondence is then used to express the partition function $Z$ as
\begin{eqnarray}
Z = \Gamma
\int d{\bf r}_a\int d{\bf r}_b \;\rho({\bf r}_a,{\bf r}_b;L)
\label{def-Z-quantum}
\end{eqnarray}
where $\Gamma$ is a normalization constant with dimension $(length)^{-2}$, whose precise value is unimportant to 
the arguments that follow\cite{Gamma},  
where ${\bf r}_a={\bf r}(0)$ and ${\bf r}_b={\bf r}(L)$ denote the endpoint positions of the flux line
and where
$\rho({\bf r}_a,{\bf r}_b;L)$ is the density matrix
\begin{equation}
\rho({\bf r}_a,{\bf r}_b;L) = \langle{\bf r}_a|\mbox{e}^{-L{\cal H}/k_BT}|{\bf r}_b\rangle
\nonumber
\end{equation}
associated with the ``quantum'' Hamiltonian
\begin{equation}
{\cal H} =  -\frac{(k_BT)^2}{2\tilde\varepsilon_1}\,\nabla_\perp^2 + V({\bf r})
\nonumber
\end{equation}
(here $\nabla_\perp=(\partial_x,\partial_y)$ is the gradient in the $(x,y)$ plane).
Using a complete set $\{ \psi_n({\bf r})\}$ of eigenfunctions of the operator ${\cal H}$ with eigenvalues $E_n$, 
the density matrix can be written in the form
\begin{eqnarray}
\rho({\bf r}_a,{\bf r}_b;L) = \sum_n \mbox{e}^{-LE_n/k_BT} \psi_n({\bf r}_a)\psi_n^*({\bf r}_b)
\label{rho-quantum}
\end{eqnarray}
Equations (\ref{def-Z-quantum}) and (\ref{rho-quantum}) lead to the following expression for the partition 
function
\begin{eqnarray}
Z = \Gamma\sum_n \mbox{e}^{-LE_n/k_BT} \int d{\bf r}_a\;\psi_n({\bf r}_a)\int d{\bf r}_b\;\psi_n^*({\bf r}_b)
\label{quantum-Z}
\end{eqnarray} 
It should be noted that, due to the appearance of temperature $T$ in the quantum Hamiltonian ${\cal H}$, both the 
eigenvalues $E_n$ and eigenfunctions $\psi_n$ of ${\cal H}$ will depend on temperature in a rather complicated 
way. At low enough temperatures however, we expect 
${\cal H}$ to possess many low lying, localized bound states. Under 
this assumption, we see that
in the limit $L\to\infty$, the above sum is dominated by the term corresponding to the ground state $E_0<0$
of the quantum Hamiltonian ${\cal H}$, and may be approximated by
\begin{eqnarray}
Z \simeq \Gamma\,\mbox{e}^{-LE_0/k_BT} \int d{\bf r}_a\;\psi_0({\bf r}_a)\int d{\bf r}_b\;\psi_0^*({\bf r}_b)
\label{ground-state-Z}
\end{eqnarray}
Taking $-E_0$ as an approximate expression for the effective pinning energy 
$U_e(T)$ per unit length of the flux line trapped in the columnar pin, simple quantum mechanical 
arguments\cite{Landau-Lifshitz} lead to to the conclusion  
that in this low-temperature regime, $U_e(T)$ is given by\cite{Nelson-Vinokur}~:
\begin{eqnarray}
U_e(T)= - E_0 \approx U_0 - \frac{c_1(k_BT)^2}{2\tilde\varepsilon_1b_0^2} \quad ,
\label{Ue-lowT-NV}
\end{eqnarray}
(here $c_1$ is a numerical constant of order unity).
We thus see that thermal fluctuations of the flux line average out the pinning potential of the columnar pin, 
reducing its strength from the bare value $U_0$ to $U_e(T)<U_0$. Defining the characteristic temperature $T^*$ 
such that $E_0(T^*)\simeq 0$, i.e.
\begin{eqnarray}
k_B T^* \simeq b_0\sqrt{\tilde\varepsilon_1(T^*)\, U_0}\quad,
\label{Tstar} 
\end{eqnarray}
it is found that 
two distinct physical regimes will emerge, depending on the value of the temperature with respect to the 
characteristic value $T^*$, namely~:

(i) for $T\ll T^*$,
$U_e(T)$ is reduced from its bare value according to
(here ${\bf u}(z)={\bf r}(z)-\langle{\bf r}(z)\rangle$ is the displacement of the flux line at height $z$ with    
respect to its average position in the $(x,y)$ plane)
\begin{eqnarray}
\langle u^2(z)\rangle \simeq b_0^2
\end{eqnarray}

(ii) for $T > T^*$, the potential well $V({\bf r})$ is shallow and $U_e(T)$ is strongly (exponentially) 
suppressed, 
\begin{eqnarray}
U_e(T)\approx \frac{1}{2}\,U_0\,\big(T/T^*\big)^2\;\mbox{e}^{-2(T/T^*)^2}
\end{eqnarray} 
and the mean square width of the flux line is now very large
\begin{eqnarray}
\langle u^2(z)\rangle \simeq b_0^2\;\mbox{e}^{2(T/T^*)^2}
\label{usqrd-T>Tstar}
\end{eqnarray}

A certain number of remarks seems to be appropriate at this point. First, it has been argued by Nelson and 
Vinokur\cite{Nelson-Vinokur} that in both regimes $T<T^*$ and $T>T^*$, the flux line will be {\em bound} to the 
columnar pin, although only weakly so when $T>T^*$. Altough this is perfectly true in the regime $T\ll T^*$ 
where we expect many localized states to exist and the physics to be dominated by these discrete, low lying 
states, at high temperatures 
$T>T^*$, the problem is more delicate and requires careful attention.
Indeed, when $T>T^*$, we might fall into a situation\cite{Landau-Lifshitz} where 
the only localized state is the ground state. 
In this case, the question as to whether the remaining, extended states (belonging to the 
continuum spectrum of the quantum Hamiltonian ${\cal H}$) give a nonnegligible contribution to the partition sum 
(\ref{quantum-Z}) seems to be a perfectly legitimate one.
If it turns out that the partition sum $Z$,
in some physical regime, is dominated by those extended states, 
then we would expect this regime to correspond to a
depinned flux line, i.e. a free line at high enough $T\gg T^*$ temperatures.

Our second remark is related to the form of the mean square width of the 
flux line when $T>T^*$. As $T$ grows, it can be seen 
from equation (\ref{usqrd-T>Tstar}) that $\langle u^2(z)\rangle$
grows without bound and, assuming for the sake of argument an infinite superconducting critical temperature $T_c$, 
at high enough temperature this last quantity migh become larger than the mean square width of a free flux 
line (here $d_\perp=d-1$ is the number of transverse dimensions)
\begin{eqnarray}
\langle u^2(z)\rangle_0 = \frac{{d_\perp}k_BT L}{12\tilde\varepsilon_1} \label{free-usqrd}
\end{eqnarray} 
This is, of course, conceptually unsatisfactory~: the mean square width $\langle u^2(z)\rangle$ of a ``bound'' 
flux line should be bounded from above by $\langle u^2(z)\rangle_0$, eq. (\ref{free-usqrd}), and, in the event 
that $\langle u^2(z)\rangle$ is to become very large, its upper limit should not 
exceed $\langle u^2(z)\rangle_0$. 
This limit should not, however, be put in by hand, but should emerge in a natural way from the 
formalism used to describe the physics.

In what follows, we construct such a formalism. 
The key point will be to try to incorporate, in an approximate way, all the states $\{\psi_n\}$ in the partition 
sum (\ref{quantum-Z}). 
This is a rather difficult task within the Schr\"odinger formulation, 
since it requires the evaluation of a great number of eigenvalues $E_n$ of the quantum Hamiltonian ${\cal H}$ and 
their associated eigenstates $\{\psi_n({\bf r})\}$. We thus have to look for alternate ways to solve our 
problem.

For the purpose of achieving such an alternate, and hopefully more accurate,
description of our system, let us rewrite the displacement vector ${\bf r}(z)$ as the sum 
\begin{eqnarray}
{\bf r}(z) = {\bf r}_0 + {\bf u}(z)
\end{eqnarray}
where ${\bf r}_0$ is the position in the $(xy)$ plane of the center of mass (CM) 
of the flux line, and ${\bf u}(z)$ is the
displacement of the flux line at height $z$ with respect to the CM position ${\bf r}_0$.
The partition function $Z$ of equation (\ref{normalized-Z}) can then be written in the form
\end{multicols}
\begin{equation}
Z = \frac{\int\!\!d{\bf r}_0\int{\cal D}{\bf u}(z)\,\exp\Big\{
- \beta\int_0^L dz\,\big[\frac{1}{2}\,\tilde\varepsilon_1\,\Big(\frac{d{\bf u}}{dz}\Big)^2  +
V\big({\bf r}_0+{\bf u}(z)\big) \big]
\Big\} }
{\int d{\bf r}_0\int{\cal D}{\bf r}(z)\;\exp\Big\{
- \beta\int_0^L dz\;\frac{1}{2}\,\tilde\varepsilon_1\,\Big(\frac{d{\bf u}}{dz}\Big)^2
\Big\}}
\label{normalized-Z-2}
\end{equation}
\begin{multicols}{2}
\noindent
where we have been
careful to separate the CM from the internal modes, writing our measure of integration $[d{\bf r}(z)]$
as $d{\bf r}_0\, [d{\bf u}(z)]$. Now let us imagine, just for the sake of argument, that we could succeed in
integrating the above partition function over the internal modes $\{{\bf u}(z)\}$ exactly. The result would be a
single integral over the CM mode\cite{Kleinert}
\vfill
\begin{equation}
Z_0 = \int d{\bf r}_0\,\,\mbox{e}^{-V_{eff}({\bf r}_0)/k_BT} \label{Z0}
\end{equation}
where $V_{eff}({\bf r}_0)$ is the effective potential experienced by the CM mode after averaging over the internal
degrees of freedom of the flux line. Since the original pinning potential $V({\bf r})$ is very localized,
we expect the effective potential $V_{eff}({\bf r})$ to be
localized as well. In fact, $|V_{eff}({\bf r}_0=0)|$ would be nothing more than the {\em exact} value of 
the effective pinning energy $U_e(T)$ per unit length of the flux line.

An exact integration over the internal modes of the flux line being impossible to achieve, we have to resort to 
approximations. An approximate but very accurate method which has been used successfully to calculate 
path integrals of the form (\ref{normalized-Z-2}) in various quantum statistical problems, is the variational 
approach of Feynman and 
Kleinert\cite{Feynman-Kleinert,Kleinert}. In this approach, one uses a Gaussian trial Hamiltonian to find an 
approximate expression for the effective potential $V_{eff}({\bf r}_0)$ experienced by the CM mode.
Applied to various quantum problems (see for example Kleinert\cite{Kleinert}, and references therein), this 
approximation gives results which are in very good agreement with numerical simulations, both at low 
and high temperatures, which is a good indication that it does actually take into account, in a rather 
accurate way, many more states in the partition sum (\ref{quantum-Z}) than the approximation 
(\ref{ground-state-Z}). In the next section we apply this variational approach to our problem.
In doing so, we shall only give the salient features of the calculation, and refer the reader interested in 
technical details to Kleinert's book\cite{Kleinert}, which gives the most complete 
and detailed presentation of the method present nowadays in the literature.

\section{Variational approximation for the effective potential}

We now introduce the following decomposition
of the diplacement vector ${\bf u}(z)$ in Fourier modes $q_n = 2n\pi/L$~: 
\begin{eqnarray} 
{\bf u}(z) = \sum_{n\neq 0} {\bf r}(q_n)\,\mbox{e}^{i q_n z}
\label{Fourier-modes-u}
\end{eqnarray}
The above decomposition is analogous to the decomposition of an imaginary-time trajectory
in Matsubara modes, which is familiar from the equivalent quantum problem\cite{Negele-Orland}, or to the 
decomposition of a polymer's internal modes into Rouse modes familiar from polymer physics\cite{Doi-Edwards}.  
Note that the $n=0$ mode is
excluded from the summation since it corresponds to the center of mass mode ${\bf r}_0$.
The Fourier coefficients ${\bf r}(q_n)$ are related to ${\bf u}(z)$ by
\begin{eqnarray}
{\bf r}(q_n) = \frac{1}{L}\int_0^L {\bf u}(z)\;\mbox{e}^{-iq_nz}
\end{eqnarray}
In the above representation, the
partition function $Z$ of equation (\ref{normalized-Z-2}) can be written in the form~:
\begin{eqnarray}
Z  & = & \int \frac{d{\bf r}_0}{\cal A}\int [d{\bf u}(z)]\;\exp\Big\{
-\beta\int_0^L \!\!dz\,\Big[\frac{1}{2}\,\tilde\varepsilon_1\Big(\frac{d\bf r}{dz}\Big)^2 +
\nonumber\\
& + & V\big({\bf r}_0+{\bf u}(z)\big)
\Big]\Big\}
\end{eqnarray}
where ${\cal A}$ is the transverse area of the system (in the plane which is perpendicular to the direction of 
the columnar pin). The integration measure $[d{\bf u}(z)]$ stands for\cite{Kleinert}~:
\begin{eqnarray}
[d{\bf u}(z)] = \prod_{n=1}^\infty\frac{d{\bf r}_{re}(q_n)\,d{\bf r}_{im}(q_n)}
{\big(\pi/L\beta\tilde\varepsilon_1 q_n^2\big)^{d_\perp}}
\end{eqnarray}
where ${\bf r}_{re}(q_n)$ and ${\bf r}_{im}(q_n)$ denote the real and imaginary parts of ${\bf r}(q_n)$,
respectively, and where the denominator comes from integrating the free path integral in the denominator of 
equation (\ref{normalized-Z-2}).
Following Feynman and Kleinert\cite{Feynman-Kleinert,Kleinert}, 
we introduce the following variational Hamiltonian~:
\begin{eqnarray}
H_1 = \frac{1}{2}\tilde\varepsilon_1\int_0^L \!\! dz\;\Big[\Big(\frac{d\bf u}{dz}\Big)^2 + \Omega^2({\bf r}_0)
\,{\bf u}^2(z)\Big] + L_1({\bf r}_0)
\label{var-H}
\end{eqnarray}
where $\Omega^2({\bf r}_0)$ and $L_1({\bf r}_0)$ are unknown functions of the CM position ${\bf r}_0$ to be 
determined by minimization of the variational free energy $F_v$ given by~:
\begin{eqnarray}
F_v & = & F_1 + \langle H - H_1 \rangle_1 \nonumber\\
& = & F_1 + \int_0^L \!\!dz\,\big[ \big\langle V\big({\bf r}(z)\big)\big\rangle_1 - 
\frac{1}{2}\,\tilde\varepsilon_1\,\big\langle\Omega^2({\bf r}_0)u^2(z)\big\rangle_1\,\big] + \nonumber\\ 
& - & \big\langle L_1({\bf r}_0)\big\rangle_1 \label{trial-F}
\end{eqnarray}
Here $\langle\cdots\rangle_1$ denotes an average taken with statistical weight $\exp(-H_1/k_BT)$.
On the other hand, $F_1=-k_BT\ln Z_1$ is the free energy associated with the trial Hamiltonian (\ref{var-H}),
with
\begin{eqnarray}
Z_1 & = & \int \frac{d{\bf r}_0}{\cal A}\int [d{\bf u}(z)]\;\exp\Big\{
-\frac{1}{T}\int_0^L dz\;\frac{1}{2}\;\tilde\varepsilon_1\;\Big[\Big(\frac{d\bf r}{dz}\Big)^2 +
\nonumber\\
& + & \Omega^2({\bf r}_0)
\big({\bf r}(z)-{\bf r}_0\big)^2\Big]
\Big\}\;\mbox{e}^{-L_1({\bf r}_0)/T}
\end{eqnarray}

We now use the Fourier decomposition (\ref{Fourier-modes-u}) to rewrite
the quantity $\beta H_1$ in the form
\begin{eqnarray}
\beta H_1 = L\beta\tilde\varepsilon_1\sum_{n=1}^\infty (q_n^2 + \Omega^2({\bf r}_0)|{\bf r}(q_n)|^2 
+ L_1({\bf r}_0)
\end{eqnarray}
We thus obtain for the trial partition function $Z_1$~:
\begin{eqnarray}
&{Z_1}& =  \int \frac{d{\bf r}_0}{\cal A}\,\mbox{e}^{-\beta L_1({\bf r}_0)}
\int[d{\bf u}(z)] \times\nonumber\\
&\times&\exp\Big\{-L\beta\tilde\varepsilon_1\sum_{n=1}^\infty(q_n^2 + \Omega^2({\bf r}_0)\big)
\,[{\bf r}_{re}^2(q_n) + {\bf r}_{im}^2(q_n)]
\Big\} \nonumber\\
& = & \int \frac{d{\bf r}_0}{\cal A}\,\mbox{e}^{-\beta L_1({\bf r}_0)}\,\Phi\big(\Omega({\bf r}_0)\big)
\label{res-Z1}
\end{eqnarray}
Here and below, we denote by $\Phi\big(\Omega({\bf r}_0)\big)$ the quantity
\begin{eqnarray}
\Phi\big(\Omega({\bf r}_0)\big) & = & 
\prod_{n=1}^\infty \Big(\frac{q_n^2}{q_n^2+\Omega^2({\bf r}_0)}\Big)^{d_\perp} \nonumber\\
& = & \Big(\frac{L\Omega({\bf r}_0)/2}{\sinh\big(L\Omega({\bf r}_0)/2\big)}\Big)^{d_\perp}
\label{def-Phi}
\end{eqnarray}
where, in going from the first to the second line, we used the well-known infinite product 
identity\cite{Prudnikov,Kleinert}
\begin{eqnarray}
\prod_{n=1}^\infty \Big(1+\frac{x^2}{n^2\pi^2}\Big) = \frac{\sinh x}{x} \nonumber
\end{eqnarray}

We now evalute the next term in the trial free energy $F_v$. We have~:
\begin{eqnarray}
\langle V\big({\bf r}(z)\big)\rangle_1 &=& Z_1^{-1}\int \frac{d{\bf r}_0}{\cal A} \;
\mbox{e}^{-\beta L_1({\bf r}_0)}\!\int \!d[{\bf u}(z)]\,\mbox{e}^{-\beta H_1}\times\nonumber\\
&\times&\int_{\bf k} V({\bf k})\,\mbox{e}^{i{\bf k}\cdot\big[{\bf r}_0+\sum_{n\neq 0}
{\bf r}(q_n)\mbox{e}^{iq_n z}\big]}
\label{avg-V}
\end{eqnarray}
where the Fourier transform $V({\bf k})$ is given by~:
\begin{eqnarray}
V({\bf k}) = \int d{\bf r}\; V({\bf r})\,\mbox{e}^{-i{\bf k}\cdot{\bf r}}
\end{eqnarray}
and where we used the shorthand notation $\int_{\bf k}\equiv\int d^{d_\perp}{\bf k}/(2\pi)^{d_\perp}$.
Writing the sum $\sum_{n\neq 0}{\bf r}(q_n)$ in the form (here and below, $c.c.$ denotes complex conjugation)~:
\begin{eqnarray}
\sum_{n\neq 0} &{\bf r}(q_n)&\mbox{e}^{iq_n z}  =  \sum_{n=1}^\infty \big({\bf r}(q_n)\mbox{e}^{iq_n z}+ 
c.c.\big)
\nonumber\\
& = & 2\sum_{n=1}^\infty \big[ {\bf r}_{re}(q_n)\cos(q_nz) - {\bf r}_{im}(q_n)\sin(q_nz)\big]
\nonumber
\end{eqnarray}
and performing the resulting Gaussian integrations in equation (\ref{avg-V}) above, we obtain~:
\begin{eqnarray}
\langle V\big({\bf r}(z)\big)\rangle_1 &=& \frac{1}{Z_1}\int \frac{d{\bf r}_0}{\cal A}\,
\mbox{e}^{-\beta L_1({\bf r}_0)}\,\Phi\big(\Omega({\bf r}_0)\big)\,\tilde V({\bf r}_0)
\end{eqnarray}
In the above expression, we defined the function
\begin{eqnarray}
\tilde V({\bf r}_0) = 
\int_{\bf k} V({\bf k})\;\mbox{e}^{-\frac{1}{2}k^2 a^2({\bf r}_0)+i{\bf k}\cdot{\bf r}_0} 
\end{eqnarray}
which can be interpreted\cite{Kleinert} as an average of the original pinning potential $V({\bf r}_0)$ over a 
region of radius $a({\bf r}_0)$ around ${\bf r}_0$, where $a({\bf r}_0)$ is given by~:
\begin{eqnarray}
a^2(&{\bf r}_0&)  =  \frac{2}{L\beta\tilde\varepsilon_1}\sum_{n=1}^\infty\frac{1}{q_n^2 + \Omega^2({\bf r}_0)}
\nonumber\\
& = & \frac{1}{L\beta\tilde\varepsilon_1\Omega^2({\bf r}_0)}\Big(
\frac{L\Omega({\bf r}_0)}{2}\coth\Big(\frac{L\Omega({\bf r}_0)}{2}\Big)-1
\Big) 
\label{def-a0}
\end{eqnarray}
where, in going from the first to the second line, we made use of the formula\cite{Prudnikov}~:
\begin{eqnarray}
\sum_{n=1}^\infty\frac{1}{n^2 + a^2} = \frac{\pi}{2a}\,\coth(\pi a) - \frac{1}{2a^2}
\end{eqnarray}

The third average in the trial free energy (\ref{trial-F}) is evaluated in a similar fashion. We have~:
\begin{eqnarray}
&\langle& \Omega^2({\bf r}_0)u^2(z)\rangle_1  =  
2\sum_{n=1}^\infty\langle\Omega^2({\bf r}_0)|{\bf r}(q_n)|^2\rangle_1 \nonumber\\
& = & \frac{1}{Z_1}\int \frac{d{\bf r}_0}{\cal A}\;\Omega^2({\bf r}_0)\mbox{e}^{-\beta L_1({\bf r}_0)}
\int [d{\bf u}(z)]
\times\nonumber\\
&\times&\exp\Big\{-L\beta\tilde\varepsilon_1\sum_{l=1}^\infty(q_l^2+\Omega^2\big({\bf r}_0)\big)
\big[{\bf r}_{re}(q_l)+ {\bf r}_{im}(q_l)\big]\Big\} \nonumber\\
&\times& 2\sum_{n=1}^\infty\big( {\bf r}_{re}^2(q_n)+ {\bf r}_{im}^2(q_n) \big)
\end{eqnarray}
Again, evaluating the Gaussian integrals over the internal modes ${\bf r}(q_n)$, we obtain~:
\begin{eqnarray}
\int_0^L &dz& \;\frac{1}{2}\tilde\varepsilon_1\;\big\langle\Omega^2({\bf r}_0)u^2(z)\big\rangle_1 = 
\frac{1}{Z_1}\int \frac{d{\bf r}_0}{\cal A}\; \mbox{e}^{-\beta L_1({\bf r}_0)} \times
\nonumber\\
&\times& 
\frac{1}{2} d_\perp L \tilde\varepsilon_1\Phi\big(\Omega({\bf r}_0)\big)\Omega^2({\bf r}_0) a^2({\bf r}_0)
\end{eqnarray}

The last average we need for the evaluation of $F_v$ is $\langle L_1({\bf r}_0)\rangle_1$, which can be easily 
obtained, and is given by~:
\begin{eqnarray}
\langle L_1({\bf r}_0)\rangle_1 = \frac{1}{Z_1}\int \frac{d{\bf r}_0}{\cal A}\;\Phi\big(\Omega({\bf r}_0)\big)
\mbox{e}^{-\beta L_1({\bf r}_0)}L_1({\bf r}_0)
\end{eqnarray}

Collecting all terms, we finally obtain for the trial free energy $F_v$ the following expression~:
\begin{eqnarray}
F_v & = & F_1 + \frac{1}{Z_1}\int \frac{d{\bf r}_0}{\cal A}\;
\Phi\big(\Omega({\bf r}_0)\big)\;\mbox{e}^{-\beta L_1({\bf r}_0)} \times
\nonumber\\
&\times& \big[L\tilde V({\bf r}_0) - \frac{1}{2} d_\perp L\tilde\varepsilon_1\Omega^2({\bf r}_0)a^2({\bf r}_0) 
- L_1({\bf r}_0)\big]
\end{eqnarray}
Variation of the above free energy with respect to $L_1({\bf r}_0)$ shows that $F_v$ is extremal, 
and is in fact minimal\cite{Feynman-Kleinert,Kleinert}, for
\begin{eqnarray}
L_1({\bf r}_0) = L\tilde V({\bf r}_0) - \frac{1}{2} d_\perp L\tilde\varepsilon_1\Omega^2({\bf r}_0)a^2({\bf r}_0)
\label{res-L1}
\end{eqnarray}
The variational free energy $F_v$ thus reduces to~:
\begin{eqnarray}
F_v = F_1 = -k_BT\ln Z_1
\end{eqnarray}
with $Z_1$ given by expression (\ref{res-Z1}), with $L_1({\bf r}_0)$ replaced by the result (\ref{res-L1}) 
above~:
\begin{eqnarray}
Z_1 & = & \int \frac{d{\bf r}_0}{\cal A}\;\Phi\big(\Omega({\bf r}_0)\big)
\mbox{e}^{-\beta[ L\tilde V({\bf r}_0) - \frac{1}{2}d_\perp L\tilde\varepsilon_1\Omega^2({\bf r}_0)
a^2({\bf r}_0)]} \nonumber
\end{eqnarray}
From this last equation, we see that $Z_v=\mbox{e}^{-\beta F_v}=\mbox{e}^{-\beta F_1}$ can be written in the form
\begin{eqnarray}
Z_v = \int \frac{d{\bf r}_0}{\cal A} \;\mbox{e}^{-\beta W({\bf r}_0)}
\end{eqnarray}
This is exactly the goal that we set out for in the introduction~: $W({\bf r}_0)$ is no more than
the effective potential experienced by the center of mass of the flux line when all the internal modes have been 
integrated out. Within our variational approximation, it is given by~:
\begin{eqnarray}
W({\bf r}_0) & = & - k_BT \ln \Phi\big(\Omega({\bf r}_0)\big) + 
L\tilde V({\bf r}_0)  +\nonumber\\
& - & \frac{1}{2}d_\perp L\tilde\varepsilon_1\Omega^2({\bf r}_0)a^2({\bf r}_0)
\label{result-W}
\end{eqnarray}
Further minimization with respect to $\Omega^2({\bf r}_0)$ leads to the result\cite{Feynman-Kleinert,Kleinert} 
that this last quantity is given by
\begin{eqnarray}
\Omega^2({\bf r}_0)  & = & \frac{2}{d_\perp\tilde\varepsilon_1}
\;\frac{\partial\tilde V({\bf r}_0)}{\partial a^2({\bf r}_0)}
\nonumber\\
& = & \frac{1}{d_\perp\tilde\varepsilon_1}\int_{\bf k} k^2V({\bf k})
\mbox{e}^{ -\frac{1}{2}k^2a^2({\bf r}_0)+i{\bf k}\cdot{\bf r}_0 }  
\label{expr-Omega}
\end{eqnarray}

Taking for the pinning potential of a columnar pin the Gaussian form
\begin{eqnarray}
V({\bf r}) = - U_0\exp\big(-\frac{r^2}{2b_0^2}\big) \label{def-V}
\end{eqnarray}
we find for $\tilde V({\bf r}_0)$ the following expression~:
\begin{eqnarray}
\tilde V({\bf r}_0) = -U_1\;\exp\Big(-\frac{r_0^2}{2\big(b_0^2+ a^2({\bf r}_0)\big)}\Big) 
\label{tilde-V}
\end{eqnarray}
where $U_1$ is given by~:
\begin{eqnarray}
U_1 = U_0\;\Big(\frac{b_0^2}{b_0^2+a^2({\bf r}_0)}\Big)^{d_\perp/2}
\label{result-Ue}
\end{eqnarray}

In the following we shall mostly be interested in the effective depth of the pinning potential, and hence we shall 
be interested only in the values of $\Omega^2({\bf r}_0)$ and $a^2({\bf r}_0)$ evaluated at ${\bf r}_0=0$.
Using the expression (\ref{def-V}) of the pinning potential in equation (\ref{expr-Omega}) and performing the 
${\bf k}$ integral, we easily obtain the following result for $\Omega({\bf r}_0=0)$
(henceforth, we shall take $d_\perp=2$)~:
\begin{eqnarray}
\Omega(0) = \sqrt{\frac{U_0}{\tilde\varepsilon_1}}\frac{b_0}{b_0^2+a^2(0)}
\label{result-Omega0}
\end{eqnarray}
At very large sample thickness $L$ and low enough temperature, the lowest lying, localized states dominate the 
partition sum (\ref{quantum-Z}). We expect in this regime a localized flux line, with finite values of the parameters 
$\Omega(0)$ and $a(0)$. Assuming that $\Omega(0)$ remains finite (nonzero), so that 
$L\Omega(0)$ is very large when $L\to \infty$, we can approximate $a^2(0)$ from equation 
(\ref{def-a0}) in the above limit by the following expression~:
\begin{eqnarray}
a^2(0) \simeq \frac{1}{2\beta\tilde\varepsilon_1\Omega(0)}\quad,\quad L\to\infty
\label{asymp-res-a0}
\end{eqnarray}
Using equation (\ref{result-Omega0}) to eliminate $\Omega(0)$ from equation (\ref{asymp-res-a0}), we obtain the 
following equation for $a^2(0)$~:
\begin{eqnarray}
a^2(0) = \frac{b_0k_BT}{2\sqrt{U_0\tilde\varepsilon_1}}\Big( 1 + \frac{a^2(0)}{b_0^2}\Big)
\end{eqnarray}
Solving the above equation for $a^2(0)$, we obtain the result~:
\begin{eqnarray}
a^2(0) = \frac{b_0k_BT}{2\sqrt{U_0\tilde\varepsilon_1}\left(1 - 
\frac{k_BT}{2b_0\sqrt{U_0\tilde\varepsilon_1}}\right)}
\label{result-asqrd0-lowT}
\end{eqnarray}
which shows that the mean square width of the flux line $a^2(0)$ {\em diverges} at the temperature 
$T_{dp}$ given by~:
\begin{eqnarray}
k_BT_{dp} = 2b_0\sqrt{U_0\,\tilde\varepsilon_1(T_{dp})}
\label{dep-T-ours}
\end{eqnarray}
Before proceeding any further, we need to observe that expression (\ref{result-asqrd0-lowT}) for $a^2(0)$ 
is valid only in the temperature range $0\le T<T_{dp}$ (for $T>T_{dp}$, the denominator in equation
(\ref{result-asqrd0-lowT}) becomes negative, which would violate the requirement $a^2(0)\ge 0$). 
We therefore expect the approximation (\ref{asymp-res-a0}), which relied on the assumption that $\Omega(0)$
was finite ({\em i.e.} nonzero), to be no longer valid when $T>T_{dp}$. Assuming, in this regime of temperatures, 
that $L\Omega(0)\ll 1$, and using the Taylor expansion of the hyperbolic cotangent near $x=0$~:
\begin{eqnarray}
\coth x = \frac{1}{x} + \frac{x}{3} + \mbox{o}(x^3)\quad,\quad x\to 0 \nonumber
\end{eqnarray}
we obtain
\begin{eqnarray}
\frac{L\Omega}{2}\coth\Big(\frac{L\Omega}{2}\Big) - 1 \simeq \frac{L^2\Omega^2(0)}{12}
\nonumber
\end{eqnarray}
Inserting this expression back into equation (\ref{def-a0}), we obtain~:
\begin{eqnarray}
\langle u^2\rangle = d_{\perp} a^2(0) = \frac{{d_\perp}k_BTL}{12\tilde\varepsilon_1}
\label{asqrd-free}
\end{eqnarray}
which is nothing more than the mean square width of a free flux line, equation (\ref{free-usqrd}).

The above results suggest that the flux line undergoes at the critical temperature $T=T_{dp}$ a transition from a 
low temperature $(T<T_{dp})$, pinned phase to a high temperature phase $(T>T_{dp})$ where the flux line is 
delocalized. 
Note the similarity of our ``depinning'' temperature $T_{dp}$, equation (\ref{dep-T-ours}), to the characteristic 
temperature $T^*$ of Nelson and Vinokur\cite{Nelson-Vinokur,Blatter-et-al}, 
equation (\ref{Tstar}); the two expressions differ only by a constant of order 
unity, which should in fact be considered good agreement, given the fact that different analytic forms for the 
pinning potential are used in the two cases. Unlike the quantum approach however, our variational method yields a 
mean square width $\langle u^2\rangle=d_{\perp}a^2(0)$, equation (\ref{asqrd-free}), which does not grow without 
bound with 
temperature as in equation (\ref{usqrd-T>Tstar}). The fact that we are able, 
in the high temperature phase, to obtain the correct expression for 
the mean square width of a free line from our equations is an indication that our 
variational approach is superior to the approximation (\ref{ground-state-Z}), and that the resulting scenario 
of a true phase transition might indeed be the correct way to describe the physics of our system.

Equations (\ref{result-W}), (\ref{result-Ue})-(\ref{result-Omega0}), 
and (\ref{dep-T-ours})-(\ref{asqrd-free}), are the 
main results of this paper. In particular, equation (\ref{result-W}) gives the effective potential experienced by 
the center of mass of the flux line after all the internal modes have been integrated out, and as such, it 
allows us to discuss the physics of 
the flux line near the columnar pin in terms of only one variable, namely the position ${\bf r}_0$ of the center 
of mass mode. This discussion will be the subject of the next section.

\section{Discussion and Conclusions}

We now can construct a complete and consistent picture of the physical behaviour of a single flux line near a 
columnar pin. As we did in the introduction, we shall consider two separate temperature regimes~:

(i) For $T<T_{dp}$, the internal fluctuations of the flux line smear out the pinning potential
only weakly, so that the 
center of mass of the flux line still experiences a finite (nonvanishing) effective potential $W({\bf r}_0)$.
We can therefore think of the center of mass mode of the flux line ${\bf r}_0$ as an ordinary classical 
particle trapped in the pinning potential $W({\bf r}_0)$, 
and in contact with a thermal reservoir at temperature $T$. 
Using equations (\ref{result-W}), (\ref{tilde-V}), (\ref{result-W}) and (\ref{result-asqrd0-lowT}), 
and the definition (\ref{def-Phi}) of the function $\Phi$, 
one can easily show that the strength $U_t(T)=|W({\bf r}_0=0)|$ of the effective 
potential $W({\bf r}_0)$ (which is also the total pinning energy of the flux line) can be written in the form 
$U_t(T)=LU_e(T)$, where the effective energy {\em per unit length} $U_e(T)$
\begin{eqnarray}
U_e(T) = U_0\Big( 1 -\frac{k_BT}{2b_0\sqrt{U_0\tilde\varepsilon_1(T)}}\Big)^2
\end{eqnarray}
is reduced with respect to the bare pinning energy per unit length $U_0$ and vanishes exactly at $T=T_{dp}$.
Since the average exit time\cite{Gardiner,Risken} $\tau_e$ 
of a classical (overdamped) particle in contact with a thermal bath at temperature $T$, from a potential well 
of depth $U_t$ is proportional to 
the Boltzmann factor $\exp(U_t/k_BT)$, we see that, in the case of the center of mass mode 
of the flux line where $\exp(U_t(T)/k_BT)=\exp(LU_e(T)/k_BT)$, no matter how small $U_e(T)$ is, in 
the limit of a large sample thickness $L$, $\tau_e$ will be very large, so that ergodicity is effectively broken 
and the flux line may be considered, for all practical purposes, as localized in the vicinity of the columnar pin.
In the boson language, this is what is meant by saying that the flux line is
``bound'' to the columnar defect. In this low temperature regime, only low lying, localized states give a 
significant contribution to the sum (\ref{quantum-Z}), and expression (\ref{ground-state-Z}) is ``good enough'' an 
approximation to the partition function.

(ii) For $T>T_{dp}$, the internal modes fluctuations diverge, and the mean square width of the flux line takes on 
its ``free'' value. As a result, the pinning potential is completely smeared out~: the effective pinning energy 
$U_t(T)=|W({\bf r}_0=0)|$ experienced by the center of mass mode vanishes
(as can easily be verified using equations (\ref{def-Phi}), (\ref{result-W}), (\ref{result-Omega0}) and 
(\ref{asqrd-free})), 
and the flux line is not pinned at all by the columnar defect.
In this regime, (if it does actually exist, and is not just an artifact of our variational approximation),
one has to be careful not to neglect the extended states (in the continuum part of the spectrum 
of the quantum Hamiltonian ${\cal H}$) in the partition sum (\ref{quantum-Z}), as these states 
turn out to dominate the high temperature behaviour. The flux line wanders freely around the columnar pin, and 
this free behaviour emerges in the most natural of ways from our equations.

In conclusion, in this paper we applied the powerful and yet simple Feynman--Kleinert variational method to 
investigate the behaviour of a flux line near a columnar pin in type II superconductors. Our results suggest that 
the flux line undergoes a phase transition from a low temperature phase where it is effectively pinned to the 
defect, to a high temperature phase where the line behaves as a free flux line. While there are some good 
indications that our theory might be more accurate than previous work\cite{Nelson-Vinokur,Blatter-et-al}, the 
issue of whether a true phase transition does indeed exist remains open.
It therefore appears that the only way to decide whether there is an actual phase transition
at $T=T_{dp}$ is to resort to other techniques such as renormalization group methods or
numerical simulations.

\acknowledgments

The author acknowledges discussions with Professor Leo Radzihovsky. 
This work was supported by the NSF through grant  DMR--9625111.

\section{Appendix A~:~ Fluctuations of a single free flux line}

In order to make this paper self-contained, we here present the derivation of the mean square width
$\langle u^2\rangle$ of a single free flux line. 
The Hamiltonian (\ref{el-H-1}) of a free flux line can be rewritten
within the Rouse decomposition (\ref{Fourier-modes-u}) in the form
\begin{eqnarray}
H =  \sum_{n\neq 0}\;\frac{1}{2}\;(L\tilde\varepsilon_1)\,\, q_n^2\,|{\bf r}(q_n)|^2
\end{eqnarray}
From this expression, we see that the free elastic propagator $G_0(q_n)$ such that
\begin{eqnarray}
\langle r_\alpha(q_n)r_\beta(q_m)\rangle = k_BTG_0(q_n)\delta_{\alpha,\beta}\delta_{n,-m}
\label{avg-qmode}
\end{eqnarray}
is given by
\begin{eqnarray}
G_0(q_n) = \frac{1}{L\tilde\varepsilon_1q_n^2} \label{propagator} 
\end{eqnarray}
so that the mean square width of the flux line is given by
\begin{eqnarray}
\langle u^2(z)\rangle & = & d_\perp k_BT\sum_{n\neq 0}^\infty G_0(q_n) \nonumber\\
& = & \frac{d_\perp k_BTL}{12\tilde\varepsilon_1}
\label{u2-free}
\end{eqnarray}
where, in going from the first to the second line, we used the fact that $\sum_{n\ge 1}n^{-2}=\pi^2/6$.

For completeness, we here also compute the average ``tipping angle'' 
$\langle(d{\bf r}/dz)^2\rangle$ of the free flux line 
as well as the relative displacement $\langle[{\bf u}(z)-{\bf u}(z')]^2\rangle$.
The average tipping angle of the line is given by
\begin{eqnarray}
\langle(d{\bf r}/dz)^2\rangle = 2d_\perp k_BT\sum_{n=1}^{N_\xi} q_n^2G_0(q_n) \label{interm-drdz}
\end{eqnarray}
where, in order to obtain a finite result, we introduced the cut-off $N_\xi\simeq L/\xi_c$, which corresponds to 
summing over all wave-vectors $q_n$ such that $|q_n|<2\pi/\xi_c$ (the factor 2 on the {\em rhs} of the above 
equation comes from the negative $q_z$ modes). 
Equation (\ref{interm-drdz}) then yields
\begin{eqnarray}
\langle(d{\bf r}/dz)^2\rangle \simeq \frac{2d_\perp k_BT}{\tilde\varepsilon_1\xi_c}
\label{result1-drdz}
\end{eqnarray}
which is basically the result quoted in Appendix B of reference \cite{Nelson-Vinokur}. In the regime of 
temperatures where the coherence length $\xi_c(T)$ along the easy axis of a layered material 
is smaller than the layer spacing $s$, we cut off the sum in equation (\ref{interm-drdz}) at 
$|q_n|<2\pi/s$, which 
gives $\langle(d{\bf r}/dz)^2\rangle = (2d_\perp k_BT/\tilde\varepsilon_1 s)$ instead of 
(\ref{result1-drdz}).

For the mean square displacement $\langle[{\bf u}(z)-{\bf u}(0)]^2\rangle$
we find~:
\begin{eqnarray}
\langle[{\bf u}(z)-{\bf u}(0)]^2\rangle = 4d_\perp k_BT\sum_{n=1}^\infty G_0(q_n)\,[1-\cos q_n(z)]
\nonumber
\end{eqnarray}
Using expression (\ref{propagator}) of the elastic propagator, and transforming the sum into an integral, we
obtain
\begin{eqnarray}
\langle[{\bf u}(z)-{\bf u}(0)]^2\rangle \simeq \frac{2d_\perp k_BT}{\pi\tilde\varepsilon_1}\int_0^\infty
dq\,\,\frac{1-\cos qz}{q^2} 
\nonumber
\end{eqnarray}
Changing the variable of integration form $q$ to $u=|z|q$, we finally obtain
\begin{eqnarray}
\langle[{\bf u}(z)-{\bf u}(0)]^2\rangle = \frac{d_\perp k_BT}{\tilde\varepsilon_1}\,|z|
\label{diff-u-free}
\end{eqnarray}
where we used the fact that\cite{Abramowitz} $\int_0^\infty du\,(1-\cos u)/u^2 =\pi/2$.

\end{multicols}
\end{document}